\documentclass{article}
\usepackage{hyperref}
\usepackage{amsmath,amsthm,amsfonts}
\usepackage{pgfplotstable}
\usepackage{authblk}

\newtheorem{definition}{Definition}[section]
\newtheorem{question}[definition]{Question}
\DeclareMathOperator{\Sp}{\textrm{Sp}}
\begin{document}

\title{Sparsification of Matrices and Compressed Sensing}
\author[1]{Fintan Hegarty}
\author[2]{Padraig \'{O} Cath\'{a}in}
\author[3]{Yunbin Zhao}
\affil[1]{School of Mathematics, Statistics, and Applied Mathematics,\newline National University of Ireland, Galway}%\newline fintan.hegarty@gmail.com}
\affil[2]{Department of Mathematical Sciences,
Worcester Polytechnic Institute}%\newline p.ocathain@wpi.edu}
\affil[3]{School of Mathematics\\
University of Birmingham}%\newline y.zhao.2@bham.ac.uk}
\affil[]{\textit{fintan.hegarty@gmail.com, $^{\text{\upshape 2}}$p.ocathain@wpi.edu, $^{\text{\upshape 3}}$y.zhao.2@bham.ac.uk}}
%\email[1]{fintan.hegarty@gmail.com}
%\email[2]{p.ocathain@wpi.edu}
%\email[3]{y.zhao.2@bham.ac.uk}
\date{}
\maketitle

%%%%%%%%%% AUTHOR: ADD YOUR ABSTRACT:
\begin{abstract}
\advance\leftskip by 6pt
\advance\rightskip by 6pt
Compressed sensing is a signal
processing technique
whereby the limits imposed by the Shannon--Nyquist theorem can be
exceeded provided certain conditions are imposed on the signal. Such
conditions occur in many real-world scenarios, and compressed
sensing has emerging applications in medical imaging, big data,
and statistics.
Finding practical matrix constructions and computationally efficient
recovery algorithms for compressed sensing is an area of intense
research interest. Many probabilistic matrix constructions have been
proposed, and it is now well known that matrices with entries drawn
from a suitable probability distribution are essentially optimal for
compressed sensing.

Potential applications have motivated the search for
constructions
of sparse compressed sensing matrices (i.e., matrices
containing few non-zero entries). Various constructions have been
proposed, and simulations suggest that their performance is comparable
to that of dense matrices. In this paper, extensive simulations
are presented which suggest that sparsification leads to a marked
improvement in compressed sensing performance for a large class of
matrix constructions and for many different recovery algorithms.
\end{abstract}

%%%%%%%%%%%%%%%%%%%%%%%  AUTHOR's MAIN TEXT BEGINS HERE:
%%%%%%%%%%%%%%%%%%%%%%%  Use amsart conventions.
%%%%%%%%%%%%%%%%%%%%%%

\section{Introduction} \label{SectionIntro}
Compressed sensing is a new paradigm in signal processing,
developed in a series of ground-breaking publications by Donoho,
Cand\`{e}s, Romberg, Tao and their collaborators over the
past ten years or so \cite{DonohoCS,CandesRombergTaoRobustUncertainty,CandesRombergTaoStableRecovery}. Many real-world signals have the
special property of being \textit{sparse} --- they can be stored much
more concisely than a random signal. Instead of sampling the whole
signal and then applying data compression algorithms, sampling and
compression of sparse signals can be achieved simultaneously. This
process requires dramatically fewer measurements than the number
dictated by the Shannon--Nyquist Theorem, but requires complex
measurements
which are
\textit{incoherent} with respect to the
signal. The compressed sensing paradigm has generated an explosion
of interest over the past few years within both the mathematical and
electrical engineering research communities.\looseness=-1

A particularly significant application has been to Magnetic Resonance
Imaging (MRI), for which compressed sensing can speed up scans by a
factor of five~\cite{LustigDonohoPauly}, either allowing increased
resolution from a given number of samples or allowing real-time
imaging at clinically useful resolutions. A major breakthrough
achieved with compressed sensing has been real-time imaging of the
heart \cite{UeckerHeartRealTime,DonohoHeartRealTime}. The US National
Institute for Biomedical Imaging and Bioengineering published a news
report in September 2014 describing compressed sensing as offering a
``vast improvement'' in paediatric MRI imaging~\cite{KernMRI}. Emerging
applications of compressed sensing in data mining and computer vision
were described by
Cand\`{e}s in a plenary lecture at the 2014 International Congress~of~Mathematicians \cite{CandesICM}.

The central problems in compressed sensing can be framed in terms
of linear algebra. In this model, a signal is a vector $v$ in some
high-dimensional vector space,~$\mathbb{R}^{N}$. The sampling process
can be described as multiplication by a specially chosen $n \times N$
matrix~$\Phi$, called the \textit{sensing matrix}. Typically we will
have $n \ll N$, so that the problem of recovering $v$ from $\Phi v$
is massively under-determined.

A vector is $k$-\textit{sparse} if it has at most $k$ non-zero
entries. The set of $k$-sparse
vectors in $\mathbb{R}^{N}$ plays the role of the set of compressible
signals in
a communication system. The problem now is to find necessary and
sufficient conditions
so that the inverse problem of finding $v$ given $\Phi$ and $\Phi v$
is efficiently solvable.

If $u$ and $v$ are distinct $k$-sparse vectors for which $\Phi u =
\Phi v$, then one of them is
not recoverable. Clearly, therefore, we require that the images of
all $k$-sparse vectors under
$\Phi$ are distinct, which is equivalent to requiring that the
null-space of $\Phi$ does not contain
any $2k$-sparse vectors. There is no known polynomial time algorithm
to certify this property.
We refer to the problem of finding the sparsest solution $\hat{x}$
to the linear system $\Phi \hat{x} = \Phi x$ as the \textit{sparse
recovery problem}. Natarajan has shown that certain instances of this
problem are NP-hard \cite{Natarajan}.

Compressed sensing (CS) can be regarded as the study of methods for
solving the sparse recovery problem
and its generalizations (e.g., sparse approximations of
non-sparse signals,
solutions in the presence of noise) in a computationally efficient
way.
Most results in CS can be characterized either as certifications
that the sparse recovery problem is solvable for a restricted class
of matrices, or as the development
of efficient computational methods for sparse recovery for some given
class of matrices.\looseness=-1

One of the most important early developments in CS was a series of
results of
Cand\`{e}s, Romberg, Tao and their collaborators. They established
fundamental constraints for
sparse recovery: one cannot hope to recover $k$-sparse signals of
length $N$ in less than $O(k\log N)$ measurements under any
circumstances\footnote{Throughout this paper we use some standard notation for asymptotics: for functions $f, g:\mathbb{N}\rightarrow \mathbb{N}$, we write that $f=O(g)$ if there exists a constant $C$, not depending on $n$, such that $f(n)<Cg(n)$ for all sufficiently large $n$. Intuitively, the function $g$ eventually dominates $f$, up to a constant factor. We say that $f=\Theta(g)$ if there exist two constants $c, C$ such that $cg(n)<f(n)<Cg(n)$ for all sufficiently large $n$. Hence, $f$ and $g$ grow at the same rate.}. (For $k = 1$, standard results from complexity theory
show that $O(\log N)$ measurements are required.) The main
tools used to prove this result are the \textit{restricted
isometry
parameters} (RIP), which measure how the sensing matrix $\Phi$
distorts the $\ell_{2}$-norm of sparse vectors. Specifically, $\Phi$
has the $\mathrm{RIP}(k, \epsilon)$ property if, for every $k$-sparse
vector $v$, the following inequalities hold:
\[ (1-\epsilon) |v|_{2}^{2} \leq |\Phi v|_{2}^{2} \leq (1+\epsilon)
|v|_{2}^{2}.
\]
Tools from Random Matrix Theory allow precise estimations of the RIP
parameters of
certain random matrices. In particular, it can be shown
that the \textit{random Gaussian ensemble}, which has
entries drawn from a standard normal distribution, is
asymptotically optimal for compressed sensing, i.e., the number of measurements required is $O(k \log N)$. A slightly weaker result is
known for the \textit{\hbox{random} Fourier ensemble}, a random
selection of rows from the discrete Fourier transform matrix
\cite{CandesRombergTaoRobustUncertainty,CandesRombergTaoStableRecovery,TaoRandomMatrixTheory}.

As well as providing examples of asymptotically optimal compressed sensing matrices,
Cand\`{e}s et al. provided an efficient recovery algorithm:
they showed that, under modest additional assumptions on the RIP
parameters of a matrix, \hbox{$\ell_{1}$-minimization} can be used for signal
recovery. Thus efficient signal recovery is possible in large systems,
making applications to real-world problems feasible.

Generation and storage of random matrices are potential
obstacles to implementations of CS. It is also difficult to design
efficient signal recovery algorithms capable of exploiting the
structure of a random matrix. For implementation in real-world
systems, it is desirable that CS constructions produce matrices
that are \textit{sparse} (possess relatively few non-zero
entries), \textit{structured}, and \textit{deterministically
constructed}. Systems with these properties can be stored implicitly,
and efficient recovery algorithms can be designed to take advantage
of their known structure. If $\Phi$ is $n \times N$ with $d$ non-zero
\hbox{entries} per column, then computing $\Phi v$ takes $\Theta(dN)$ operations,
which is a significant saving when $d \ll n$. In some applications,
signals are frequently subject to rank-one updates (i.e., $v$
is replaced by $v+\alpha e_{i}$), in which case the image vector can
be updated in time~$O(d)$,
see \cite{ZhuGelashviliRazenshteyn}.\looseness=-1

Motivated by real-world applications, a number of papers have
explored CS constructions
where the Gaussian ensemble is replaced
by a
sparse
random matrix (e.g., coming from an expander graph
or an LDPC code) \cite{BerindeGilbertIndykKarloffStrauss,SarvothamBaronBaraniuk,Raginsky}, or by a matrix
obtained from a deterministic construction \cite{deVore,Mixon,GilbertLiPoratStrauss}. But to date, constructions meeting all
three criteria have either been asymptotic in nature (i.e.,
the results only produce matrices that are too large for practical
implementations), or are known only to exist for a very restricted
range of parameters. This investigation was inspired by work
of the second author on constructions of sparse CS matrices
from pairwise balanced designs and complex Hadamard matrices
\cite{mypaper-PBD,mypaper-CScomp}. Some related work on constructing
CS matrices from finite geometry is contained in~\cite{LiGe,XiaLiu}.

In this paper we take a new approach. Rather than constructing a
sparse matrix and examining its CS properties, we begin with a matrix
which is known to possess good CS properties (with high probability)
and explore the effect of sparsification on this matrix. That is, we
set many of the entries in the original matrix to zero, and compare
the performance of the sparse matrix with the original. Results of
Guo, Baron and Shamai suggest that sparse matrices should
behave similarly
to dense matrices in our regime \cite{GuoBaronShamai}. \hbox{Surprisingly,}
we actually observe an \textbf{improvement} in signal recovery as the
sparsity
increases.

First we survey some previous work on sparse compressed sensing
matrices. Then in Section \ref{SectionMethod}, we give a formal
definition of sparsification, and describe algorithms
used to generate random matrices and random vectors, as well as the
recovery algorithms. In Section~\ref{SectionResults} we describe the results of extensive
simulations. These provide substantial computational evidence which
suggests that sparsification is a robust phenomenon, providing benefits
in both recovery time and proportion of successful recoveries for
a wide range of random and structured matrices occurring in the CS
literature. In particular,
Table 2 shows the benefits of sparsification for a range
of matrix constructions, while Figure \ref{EffectOfSparsification}
illustrates how sparsification improves performance for a range of
CS recovery algorithms. Finally,
in Section \ref{SectionConclusion}
we conclude with some observations and open questions motivated by
our numerical experiments.

\section{Tradeoffs between sparsity and compressed \hbox{sensing}}
A number of authors have investigated ways of replacing random
ensembles with more
computationally tractable sensing matrices. As previously mentioned,
foundational results of Cand\`{e}s et al. establish asymptotically
sharp results: to recover signals of length $N$ with $k$ non-zero
entries, $n = \Theta(k\log N)$ measurements are necessary. Work of
Chandar established that when $n = \Theta(k\log N)$, then the columns
of $\Phi$ must contain at least $\Theta(\min\{k, N/n\})$ non-zero
entries \cite{Chandar}. In \cite{NelsonNguyen}, \hbox{Nelson} and Nguyen establish
an essentially optimal result when $n = \Theta(k\log N)$ and $k < N/
\log^{3} N$. They show that each column of $\Phi$ necessarily contains
$\Theta(k \log N)$ non-zero entries; i.e., the proportion of
non-zero entries in $\Phi$ cannot tend to zero as $N$ tends to $\infty$.

Observe that some restriction on $k$ as a function of $N$ is necessary:
in the limiting case $k = N$, the identity matrix clearly suffices
as a sparse sensing matrix. Furthermore, combinatorial constructions
of sparse matrices are known which have near optimal recovery guarantees with a mutual incoherence property\footnote{Informally, the $k$-RIP property requires that $k$-sets of columns of $\Phi$ approximate an orthonormal basis.
The incoherence of a matrix is the maximal inner product of a pair of columns, which is essentially the 2-RIP of $\Phi$. In contrast to $k$-RIP, efficient constructions of matrices with near-optimal incoherence are known \cite{mypaper-PBD}, but they have sub-optimal compressed sensing performance. Using 2-RIP alone, asymptotically one requires at least $k^{2}$ measurements to recover $k$-sparse signals. No practical deterministic construction is known which uses asymptotically fewer measurements; this obstruction is known as the \textit{square-root bottleneck}, and finding more efficient deterministic constructions is a major open problem in compressed sensing.}
\cite{mypaper-PBD}. In such matrices $n = O(k^{2})$,
and for certain \hbox{infinite} families of matrices (e.g., those coming
from projective planes) the number of non-zero entries in each column
is $\Theta(k)$. Results bounding errors in the $\ell_{1}$ norm (so-called
RIP-1 guarantees) have been obtained using expander graphs. In
\hbox{particular}, Bah and Tanner have shown that essentially optimal RIP-1
recovery can be achieved when $\lim_{n\rightarrow\infty} N/n = \alpha$
for some fixed $\alpha$, with a constant number of non-zero entries per
column \cite{BahTanner}. (See also the discussion of dense versus sparse
matrices in Section \ref{SectionMethod} of this paper.) These bounds are strictly
weaker than RIP-2 bounds, though fast specialised algorithms have been
developed for signal recovery with such matrices \cite{Ravanmehr}.\looseness=-1

Since the $k$-RIP property is difficult to establish in
practice, some authors have relaxed this in various
directions. Berinde, Gilbert, Indyk, Karloff and Strauss
\cite{BerindeGilbertIndykKarloffStrauss} considered random binary
matrices with constant column sum and related these to the incidence
matrices of expander graphs. We reinterpret these matrices as
sparsifications of the all-ones matrix below. Sarvotham, Baron and Baraniuk \cite{SarvothamBaronBaraniuk}
and Dimakis, Smarandache and Vontobel \cite{Dimakis}
have considered the use of LDPC matrices. In particular, they
have provided a strong correspondence between error-correcting
performance of LDPC codes (when considered over $\mathbb{F}_{2}$)
and CS performance of the same binary matrices (when considered
over $\mathbb{R}$).  While both groups obtained essentially optimal
CS performance guarantees, their constructions are limited by the
lack of known explicit constructions for expander graphs and LDPC
codes respectively. Moghadam and Radha have previously considered a two
step construction of sparse random matrices, involving construction
of a random \hbox{$(0,1)$-matrix} followed by replacing each entry 1 with
a sample from some probability distribution,~\cite{MoghadamRadha2,MoghadamRadha}.

If one is content with recovery of each sparse vector with high
probability, then much sparser matrices become useful. A strong
result in this direction is due to Gilbert, Li, Porat and Strauss,
who show that there exist matrices
with $n = O(k\log N)$ rows and
$\Theta(\log^2 k \log N)$ non-zero entries per column which recover sparse
vectors with probability
0.75 \cite{GilbertLiPoratStrauss} (see also
\cite{Tehrani}). Their matrices also come with efficient encoding,
updating and recovery algorithms. While essentially optimal results
are known for sparsity bounds on CS matrices with an optimal number of
rows, much less is known when either some redundant rows are allowed
in the construction, or when RIP is replaced with a slightly weaker
condition.

Several authors have compared the performance of sparse and dense
CS matrices \cite{GuoBaronShamai,WangWainwright,DoGan,LuKpalma}.
Guo, Baron and Shamai have \hbox{essentially} shown that in certain limiting
cases of the sparse recovery problem, dense and sparse sensing
matrices behave in a surprisingly similar manner. In particular, they
consider a variant of the recovery problem: given $\Phi$ and $\Phi x$,
what can one say about any single component of $x$? They show that, as
the size of the system becomes large (in a suitably controlled way),
the problem of estimating $x_{i}$ becomes independent of estimating~
$x_{j}$. In fact, the problem is equivalent to recovering a single
measurement of $x_{i}$ contaminated by additive Gaussian noise. They
also apply their philosophy to sparse matrices, where as the size of
the matrix becomes large, estimation of \textbf{all} signal components
becomes independent, and each can be recovered independently. We refer
the reader to the original paper for technical details \cite{GuoBaronShamai}. As a result,
under their assumptions, there should be no essential difference
between CS performance in the sparse and dense cases.

For any compressed sensing matrix $\Phi$, denote by $\delta(\Phi)$ the proportion of non-zero entries in $\Phi$. Suppose that the number of columns of $\Phi$ is a linear function of the number of rows, i.e., that $N=\alpha n$ for some fixed $\alpha\in\mathbb{R}$. Suppose further that $\lim_{n\rightarrow\infty}\delta(\Phi)n^{1-\epsilon}=0$ for any $\epsilon>0$, but that $\lim_{n\rightarrow\infty}\delta(\Phi)n$ diverges. (Consider $\delta(\Phi)=\log n/n$ for example.) Under these hypotheses, Guo, Baron and Shamai claim that sparse and dense matrices exhibit identical CS performance.
In particular, there exist matrices with $k\log k$ rows and $\Theta(\log k)$ non-zero entries per row which recover vectors of sparsity $O(k)$.
%In particular, there exist matrices with density $\unfrac{n}{\log n}(\Phi)$ which recover vectors of sparsity $\Theta(n/\log n)$.

This appears to be in conflict with Nelson and Nguyen's result, which requires
at least $\Theta(n \log\log n /\log n)$ non-zero entries in such a matrix.
The difference is that Nelson and Nguyen's result holds only when $n = O(k \log^{3} N)$, whereas Guo, Baron and Shamai consider the case $n = \Theta(N)$.

Wang,
Wainwright and Ramchandran's analysis of the number of measurements required for signal recovery depends on the
quantity $(1-\delta(\Phi))k$, which can be considered a measure of how much
information about $x$ is captured in each co-ordinate of $\Phi x$.
They show that if $(1-\delta(\Phi))k \rightarrow \infty$ as $N\rightarrow \infty$ (which corresponds
to relatively dense matrices, where, in the limit, each component of the signal is sampled infinitely often), then sparsification has no effect on
recovery, while~if $(1-\delta(\Phi))k$ remains bounded (so each component of the signal is sampled only finitely many times in expectation), then what the authors term ``dramatically more measurements'' are required. We refer the
reader to the original paper for more details \cite{WangWainwright}.\looseness=-1

\hspace{-2pt}Our simulations are close in spirit to those considered by
Guo, Baron and Shamai.
\noindent Our computations are rather surprising as they
suggest a modest \textbf{improvement} in signal recovery as we
apply a sparsifying process to certain families of CS matrices. This
improvement seems to persist across different recovery algorithms
and different matrix constructions, and does not appear to have
been noted in any of the work discussed in this section. (Though
Lu, Li, Kpalma and Ronsin have observed some improvement in CS performance
for sparse binary matrices \cite{LuLiKpalma}.) We also observed
a substantial improvement in the running times of the recovery
algorithms, which may be of interest in practical
applications.

\section{Sparsification} \label{SectionMethod}

We begin with a formal definition of sparsification.

\begin{definition}
The matrix $\Phi'$ is a \textit{sparsification} of $\Phi$ if
$\Phi'_{i,j} = \Phi_{i,j}$ for every non-zero entry of $\Phi'$. The
\textit{density} of $\Phi$,
denoted $\delta(\Phi)$, is the proportion
of non-zero entries that it contains, and the \textit{relative
density} of $\Phi'$ is the ratio $\delta(\Phi')/\delta(\Phi)$. We
write $\Sp(\Phi,s)$ for the set of all sparsifications of $\Phi$
of relative density $s$.
\end{definition}

In general, we have that $\Sp(\Phi, 1) =\Phi$, and that $\Sp(\Phi,
0)$ is the zero matrix. We also have a transitive property: if $\Phi'
\in \Sp(\Phi, s_{1})$ and $\Phi'' \in \Sp(\Phi', s_{2})$ then $\Phi''
\in \Sp(\Phi, s_{1}s_{2})$. Two independent sparsifications will not
in general be comparable: there is a partial ordering on the set
of sparsifications of a matrix, but not a total order.\looseness=-1

We illustrate our notation. Consider a Bernoulli random variable which
takes value $1$ with probability $p$ and value $0$ with probability
$1-p$ and let $\Phi$ be an $n \times N$ matrix with entries drawn from
this distribution; in short, a \textit{Bernoulli ensemble} with expected
value $p$. Then the expected density of $\Phi$ is $p$.
Writing $J$
for the all-ones matrix,
a randomly chosen $\Phi' \in \Sp(J,p)$ will also have density $p$,
and can be considered a good approximation of a Bernoulli matrix.
%we have $\Phi \in \Sp(J, p)$.
If $\Phi''$ is an independent
random sparsification (i.e., all non-zero entries of the matrix
have an equal probability of being set to zero) of $\Phi'$ with relative
density $p'$, then $\Phi''$ approximates the Bernoulli ensemble
with expected value~$pp'$. So we have both $\Phi' \in \Sp(\Phi', p')$
and $\Phi' \in \Sp(J, pp')$. Later, we will consider successive sparsifications where we begin with a dense matrix whose entries are drawn from, e.g., a normal distribution.

Bernoulli ensembles have previously been considered in the compressed sensing
literature, see \cite{RauhutRedundant} for example, though note that
the matrices here take values in $\{0,1\}$, not $\{\pm 1\}$. Such
$\{\pm 1\}$-matrices are an affine transformation of ours: $M' = 2M-J$;
as a result, the compressed sensing performance of either matrix is essentially the same.

In this paper, we will mostly be interested in \textit{pseudo-random}
sparsifications of an $n \times N$ compressed sensing matrix
$\Phi$. Specifically, for $s = t/n$, we obtain a matrix $\Phi' \in
\Sp(\Phi, s)$ by generating a pseudo-random $\{0,1\}$-matrix $S$ with
$sn$ randomly located ones per column, and returning the entry-wise
product $\Phi'=\Phi\ast S$. We will generally re-normalize $\Phi'$
so that every column has unit $\ell_{2}$-norm.

Given a matrix $\Phi$, we test its CS performance by running
simulations. Since many different methodologies occur in the
literature, we specify ours here.

Our $k$-sparse vectors always contain exactly $k$ non-zero entries,
in positions
chosen uniformly at random from the $\binom{N}{k}$ possible supports
of this size.
The entries, unless otherwise specified, are drawn from a uniform
distribution on the open interval $(0,1)$.
The vector is then scaled to have unit $\ell_{2}$-norm. Simulations
where the non-zero
entries were drawn from the absolute value of a Gaussian distribution
produced similar results.
Note that many authors use $(0,1)$- or $(0,\pm 1)$-vectors for their
simulations.
Appropriate combinations of matrices and algorithms often exhibit
dramatic improvements
of performance on this restricted set of signals.

We recover signals using $\ell_{1}$-minimization. In this paper we
will use the
\texttt{matlab} LP-solver and the implementations of \emph{Orthogonal
Matching Pursuit} (OMP) and \emph{Compressive Sampling Matching
Pursuit} (CoSaMP) algorithms which were developed by Needell and Tropp~
\cite{COSAMP}. Specifically, given a matrix $\Phi$ and
signal vector~$x$, we compute $y = \Phi x$, and solve the
$\ell_{1}$-minimization problem $\Phi \hat{x} = y$ for~$\hat{x}$.
The objective function is the $\ell_{1}$-norm of $\hat{x}$ and it is
assumed that all variables are non-negative. We consider the recovery
successful if $|x - \hat{x}|_{1} \leq c$ for some constant $c$. We
take $c = 10^{-6}$ in all the simulations presented in this paper.

\begin{figure}[t]
\begin{center}
\begin{tikzpicture}
\begin{axis}[legend pos=south west, xmin=0, xmax=60, ymin=1, width=4.5in, height=2.3in,
xlabel=Sparsity of signal vector, ylabel=\% signal vectors recovered]
\addplot [thick,blue,dashed] table [y=$0.05$, x=k]{GraphOneData.txt};
\addlegendentry{$\Sp(\Phi,0.05)$}
\addplot [ultra thick,red,dotted] table [y=$0.1$, x=k]{GraphOneData.txt};
\addlegendentry{$\Sp(\Phi,0.1)\phantom{0}$}
\addplot [black] table [y=$1$,x=k]{GraphOneData.txt};
\addlegendentry{$\Sp(\Phi,1)\phantom{0.0}$}
\end{axis}
\end{tikzpicture}
\vskip-6pt
%\advance\captionindent5pt
\caption{Effect of sparsification on signal recovery.}
\vskip-14pt
 \label{Figureon}
\end{center}
 \end{figure}
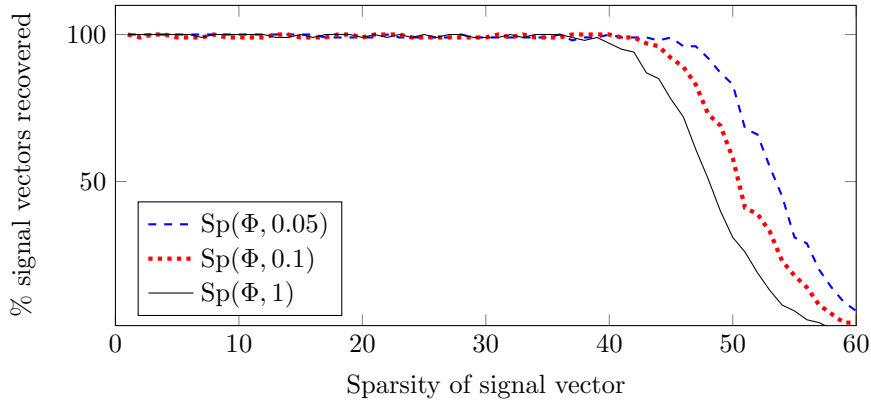

We conclude this section with an example illustrating the potential
benefits of sparsification.
In Figure \ref{Figureon}, we explore the effect of sparsification on
a $200 \times 2000$
matrix $\Phi$ with entries uniformly distributed on $(0, 1)$. The
results for this case were compared with matrices drawn from
$\textrm{Sp}(\Phi, 0.1)$ and $\textrm{Sp}(\Phi, 0.05)$. For each signal
sparsity between $1$ and $60$, we generated $500$ random vectors as
described above and recorded the number of successful recoveries
using the \texttt{matlab} LP-solver. To avoid bias we generated a
new random matrix for each trial.

We observe that for signal vector sparsities between $45$ and $55$,
matrices in $\Sp(\Phi, 0.05)$ achieve substantially
better recovery than those from $\Sp(\Phi, 1)$. The code
used to generate this simulation as well as others in this
paper is available in full, along with data from multiple
simulations at a webpage dedicated to this
project:
\texttt{http://fintanhegarty.com/compressed{\_}sensing.html} .

\section{Results} \label{SectionResults}
Our simulations produce large volumes of data. To highlight the
interesting features of these data-sets, we propose the following
measure for acceptable signal recovery in practice.

\begin{definition}
For a matrix $\Phi$ and for $0 \leq t \leq 1$, we define the
$t$-\textit{recovery threshold}, denoted $R_{t}$, to be the largest
value of $k$ for which $\Phi$ recovers $k$-sparse signal vectors with
probability exceeding $t$.
\end{definition}

We construct an estimate $\hat{R}_{t}$ for $R_{t}$ by running
simulations. As the number of simulations that we run increases,
$\hat{R}_{t}$ converges to $R_{t}$. In practice this convergence is
rapid. The definition of $R_{t}$ generalizes naturally to a space
of matrices (say $n \times N$ Gaussian ensembles): it is simply the
expected value of $R_{t}$ for a matrix chosen uniformly at random
from the space. To estimate $R_{t}$ with reasonably high confidence,
we proceed as follows: beginning with signals of sparsity $k= 1$, we
attempt 50 recoveries. We increment the value $k$ by $1$ and repeat
until we reach the first sparsity $k_{0}$ where less than $50t$
signals are recovered. Beginning at $k_{0}-3$, we attempt $200$
recoveries at each signal sparsity. When we reach a signal sparsity
$k_{1}$ where less than $200t$ signals are recovered, we attempt $1000$
signal recoveries at each signal sparsity starting at $k_{1}-3$. When
we reach a signal sparsity $k_{2}$ where less than $1000t$ signals
are recovered, we set $\hat{R}_{t} = k_{2}-1$.

We typically find that $k_{1} = k_{2}$, which gives us confidence
that $\hat{R}_{t} =\nobreak R_{t}$. Unless otherwise specified, we use the
assumptions outlined in Section \ref{SectionMethod}.

\subsection{Recovery algorithms with sparsification}
\label{SparsSec}
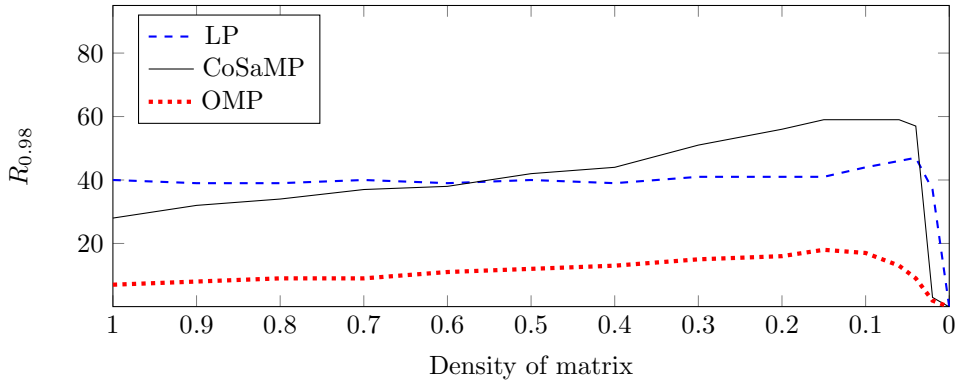
\begin{figure}
\hskip40pt\begin{tikzpicture}
\begin{axis}[legend pos=north west, xmax=1, xmin=0, ymin=0.1, ymax=95, x dir=reverse, width=5in, height=2.2in,
xlabel=Density of matrix, ylabel=$R_{0.98}$]
\addplot [dashed,thick,blue] table [y=LP, x=r]{GraphTwoData.txt};
\addlegendentry{LP\phantom{OCsa}}
\addplot [black] table [y=CoSaMP,x=r]{GraphTwoData.txt};
\addlegendentry{CoSaMP}
\addplot [dotted,ultra thick,red] table [y=OMP, x=r]{GraphTwoData.txt};
\addlegendentry{OMP\phantom{Csa}}
\end{axis}
\end{tikzpicture}
\vskip-6pt
%\advance\captionindent15pt
\caption{Signal recovery as a function of matrix density for LP,
OMP and CoSaMP.}
\label{EffectOfSparsification}
\vskip-8pt
\end{figure}

As suggested already in Figure \ref{Figureon}, taking $\Phi'
\in \Sp(\Phi, s)$ for some value of $s \sim 0.05$ seems to offer
considerable improvements when using linear programming for signal
recovery. Similar results hold for OMP and CoSaMP, though note
that in each case we supply these algorithms with the sparsity of
the signal vector. (While there is an option to withhold this data,
the recovery performance of CoSaMP seems to suffer substantially
without~it~--- and we wish to be able to perform comparisons with
linear programming.) In Figure \ref{EffectOfSparsification}, we graph
$R_{0.98}$ of $\Sp(\Phi, s)$ as a function of $s$, where $\Phi$ is a
$200 \times 2000$ matrix with entries drawn from the absolute values
of samples from a standard normal distribution.

For each algorithm, $R_{0.98}$ appears to obtain a maximum for matrices
of density between
$0.15$ and $0.05$. It is perhaps interesting to
note that the percentage improvement obtained by CoSaMP is far greater
than that for either of the other algorithms.

\begin{table}[b]
\centering
\small{
\label{Tabon}
     \begin{tabular}{|c|c  c | c  c || c  c | c  c |}
\hline
       & \multicolumn{4}{|c||}{Time for 100 recovery attempts}
       & \multicolumn{4}{|c|}{\% vectors successfully recovered}\\
     k & \multicolumn{2}{|c|}{\;CoSaMP}& \multicolumn{2}{|c||}{\;\;LP}
     & \multicolumn{2}{|c|}{\;CoSaMP}& \multicolumn{2}{|c|}{\;\;LP}\\
		 & $\delta=0.1$ & $\delta=1$ & $\delta=0.1$ &
		 $\delta=1$ & $\delta=0.1$ & $\delta=1$ & $\delta=0.1$
		 & $\delta=1$\\\hline
		    1 & 0.94 & 0.44 & 18.1 & 106.61 & 100 & 100 &
		    100 & 100\\
				10 & 0.78 & 1.25 & 39.78 & 157.53 &
				100 & 100 & 100 & 100\\
				20 & 0.56 & 1.98 & 27.50 & 177.17 &
				100 & 100 & 100 & 100\\
				30 & 1.56 & 4.48 & 27.39 & 171.84 &
				100 & 99 & 100 & 100\\
				40 & 3.68 & 33.98 & 27.02 & 207.22 &
				100 & 55 & 99 & 94\\
				50 & 11.09 & 66.77 & 33.59 & 375.17 &
				99 & 7 & 78 & 38\\
				60 & 48.38 & 81.54 & 43.82 & 364.25 &
				75 & 0 & 1 & 1\\
				70 & 89.87 & 93.62 & 41.03 & 329.22 &
				0 & 0 & 0 & 0\\\hline
     \end{tabular}\!\!
\vskip8pt
\caption{Effect of sparsification on recovery time.}
\vskip-18pt
}
\end{table}

Table 1 shows the average time taken for one hundred vector
recovery attempts using $200 \times 2000$ measurement matrices with
entries drawn from the absolute values of samples from a normal
distribution, over a range of vector sparsities. We note an
improvement in running time of an order of magnitude for linear
programming when using sparsified matrices, and an improvement when
using CoSaMP.

\subsection{Matrix constructions under sparsification}
\label{ConstSec}

In this section we explore the effect of sparsification on a number
of different constructions
proposed for CS matrices. We have already encountered the Gaussian,
Uniform and Bernoulli ensembles. We will also consider some
\textit{structured random matrices}, which still have entries drawn
from a probability distribution,
but the matrix entries are no longer independent. The \textit{partial
circulant ensemble} \cite{RauhutRombergTropp}
consists of rows sampled randomly from a circulant matrix, the first
row of which contains entries drawn uniformly
at random from some suitable probability distribution. Table 2
compares $R_{0.98}$ for $\Sp(\Phi,1)$
and $\Sp(\Phi, 0.05)$ for $200 \times 2000$ matrices from each of
the classes listed. Note that in the case of the Bernoulli ensemble,
we actually compare $\Sp(J_{200,2000}, 0.5)$ with $\Sp(J_{200,2000},
0.05)$, where $J_{200,2000}$ is an all-ones matrix. The entries of
the partial circulant matrix were drawn from a normal distribution.

We denote by $\hat{k}$ the signal sparsity $k$ for which the greatest
difference in recovery between $\Phi$ and $\Phi' \in \Sp(\Phi, 0.05)$
occurs.

\begin{table}[h]
\centering
\small{
    \begin{tabular}{| l | c  c | c  c  c|}\hline
    \text{Construction} & \multicolumn{2}{c|}{$R_{0.98}$} &
    \multicolumn{3}{c|}{\textrm{Maximal performance difference}}\\
     & $\delta=1$ & $\delta=0.05$ &$\;\;\;\;\;\;\hat{k}\;\;\;\;\;\;$
     & $\;\;\;\;\;\;\;\delta=1\;\;\;\;\;\;\;$ & $\delta = 0.05$
     \\\hline
    \text{Normal}&39&46&51&25&81\\
    \text{Uniform}&39&45&51&24&73\\
    \text{Bernoulli}&39&42&49&38&67\\
    \text{Partial Circulant}&39&46&52&22&76\\\hline
    \end{tabular}
\vskip8pt
\label{Tabtw}
    \caption{Benefit of sparsification for different matrix
    constructions.}
}
\end{table}

\subsection{Varying the matrix parameters}
\label{DimsSec}

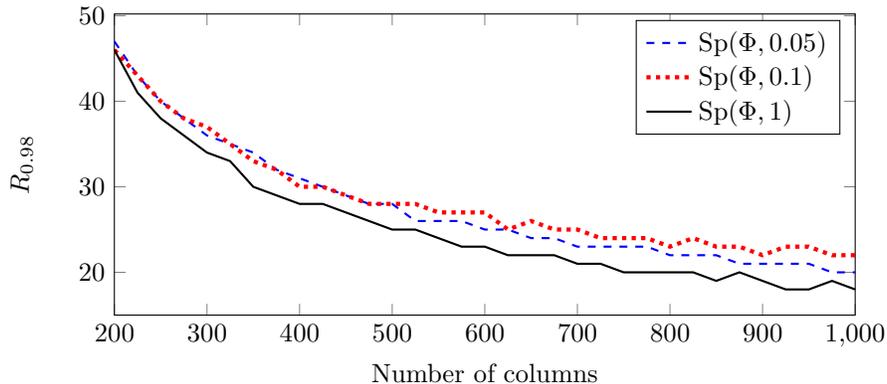
\begin{figure}[h]
\centering
\begin{tikzpicture}
\begin{axis}[xmin=200, xmax=1050, ymin=15, xmax=1000, width=4.5in, height=2.2in,
xlabel=Number of columns, ylabel=$R_{0.98}$]
\addplot [thick,dashed,blue] table [y=$0.05$, x=r]{GraphThreeData.dat};
\addlegendentry{$\Sp(\Phi,0.05)$}
\addplot [ultra thick,dotted,red] table [y=$0.1$, x=r]{GraphThreeData.dat};
\addlegendentry{$\Sp(\Phi,0.1)\phantom{0}$}
\addplot [thick,black] table [y=$1$,x=r]{GraphThreeData.dat};
\addlegendentry{$\Sp(\Phi,1)\phantom{0.0}$}
\end{axis}
\end{tikzpicture}
%\advance\captionindent-5pt
\caption{Recovery capability of matrices with 100 rows and varying
number of columns under sparsification.}
\label{figthree}
%\label{FixedRowsVaryingRatio}
 \end{figure}

Finally, we investigate the effect of sparsification on matrices of
varying parameters. In particular,
we explore the effect of sparsification on a family of matrices with
entries drawn from the absolute
value of the Gaussian distribution. First we explore the effect of
sparsification as the ratio of
columns to rows in the sensing matrix increases. For Figure \ref{figthree},
we use signal vectors whose entries were drawn from the absolute
value of the normal distribution. We observe a modest improvement in
performance which appears to persist.

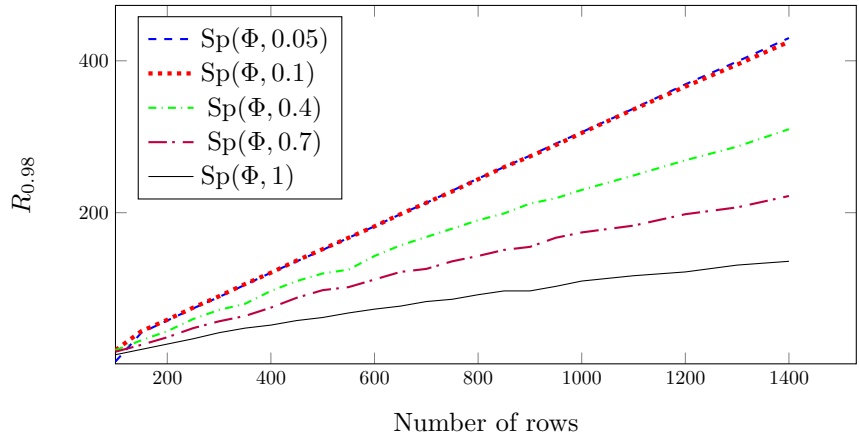
\begin{figure}[h]
\centering
\begin{tikzpicture}
\begin{axis}[every tick label/.append style={scale=0.75}, /pgf/number format/1000 sep={}, xmin=100, ymin=1, width=4.5in, height=2.5in, legend pos=north west,
xlabel=Number of rows, ylabel=$R_{0.98}$]
\addplot [thick,dashed,blue] table [y=$0.05$, x=rows]{GraphFourData.txt};
\addlegendentry{$\Sp(\Phi,0.05)$}
\addplot [ultra thick, dotted,red] table [y=$0.1$, x=rows]{GraphFourData.txt};
\addlegendentry{$\Sp(\Phi,0.1)\phantom{0}$}
\addplot [dash dot,thick,green] table [y=$0.4$, x=rows]{GraphFourData.txt};
\addlegendentry{$\Sp(\Phi,0.4)$}
\addplot [thick,dash pattern={on 7pt off 2pton 1pt off 3pt},purple] table [y=$0.7$, x=rows]{GraphFourData.txt};
\addlegendentry{$\Sp(\Phi,0.7)$}
\addplot [black] table [y=$1$,x=rows]{GraphFourData.txt};
\addlegendentry{$\Sp(\Phi,1)\phantom{0.0}$}
\end{axis}
\end{tikzpicture}
\vskip-3pt
%\advance\captionindent-5pt
\caption{Recovery with CoSaMP for matrices with fixed row to column
ratio under sparsification.}
\label{figfour}
%\label{FixedRatioVaryingRows}
 \end{figure}

Now for Figure \ref{figfour}, we fix the ratio of columns to rows
of $\Phi$ to be $10$, and vary the number of rows. We know from the
results of Cand\`{e}s et al. that $R_{0.98} = \Theta(n/\log n)$ in
all cases. Nevertheless, the clear difference in slopes for recovery
at different sparsities offers compelling evidence that the benefits
of sparsification persist for large matrices.

\section{Conclusion} \label{SectionConclusion}

Some of the most important open problems in compressed sensing relate
to the development of efficient matrix constructions and effective
algorithms for sparse recovery. Deterministic constructions are
essentially limited by the Welch bound: using known methods it is
not possible to guarantee recovery of vectors of sparsity exceeding
$\Theta(\sqrt{n})$, where $n$ is the number of rows in the recovery matrix
(see,~e.g.,~\cite{mypaper-PBD}). Probabilistic constructions
are much better: Cand\`{e}s and Tao's theory of restricted isometry
parameters allows the provable recovery of vectors of sparsity $k$ in
dimension $N$ with $\Theta(k \log(N))$ measurements. Such guarantees hold
with overwhelming probability for Gaussian ensembles and many other
classes of random matrices. But the random nature of these matrices
can make the design of efficient recovery algorithms difficult. In
this paper we have demonstrated that sparsification offers potential
improvements for computational compressed sensing. In particular,
Figures \ref{Figureon} and \ref{figfour} show
that sparsification results in the recovery of vectors of higher
sparsity. Table 1 shows a substantial improvement in runtimes
for linear programming arising from sparsification. These appear
to be robust phenomena, which persist under a variety of recovery
algorithms and matrix constructions. At the problem sizes that we
explored, matrices with densities between $0.05$ and $0.1$ seemed to
provide optimal performance.

We conclude with a small number of observations and conjectures which
we believe to be suitable for further investigation. Since a Bernoulli
ensemble in our terminology can be regarded as a sparsification of
the all-ones matrix, it is clear that sparsification can improve CS
performance. The necessary decay in CS performance as the density
approaches zero shows that the effect of sparsification cannot
be monotone. Extensive simulations suggest that when recovery is
achieved with a general purpose linear programming solver, matrices
with approximately~$10\%$ non-zero entries have substantially better
CS properties than dense matrices. A catastrophic decay of compressed
sensing performance occurs in many of the examples we investigated
between densities of $0.05$ and $0.01$. We pose two questions which
we think suitable for further research.

\begin{question}	
 %As the number of rows in
%	$\Phi$ increases, the optimal matrix density appears to
%	decrease. \textbf{Is there a function $\Gamma(n,k,N)$ of
%	the matrix parameters which describes the optimal level of
%	sparsification?} We propose that  the \redden{asymptotics of $\Gamma$
%	may be asymptotically independent}
%\marginpar{is there a better way to phrase this?}
% of the matrix construction
%	and of the recovery algorithm. We conjecture that when
%	$N < n^{\alpha}$ the optimal density of a CS matrix
%	will be approximately $\alpha n^{-\unfrac{1}{2}}$ when $k =
%	o(n^{1-\epsilon})$.
As the number of rows of $\Phi$ increases, the optimal matrix density appears to decrease. This effect does not appear to depend strongly on the matrix construction chosen. Does there exist a function $\Gamma(n,N,k)$ which describes the optimal level of sparsification for an $n\times N$ matrix recovering $k$-sparse vectors? Our simulations suggest that when $N<n^{\alpha}$, the optimal density of a CS matrix will be approximately $\alpha n^{-1/2}$ when $k=o(n^{1-\epsilon})$.
\end{question}

\begin{question}
    We have considered pseudo-random
    sparsifications in this paper. In general, this should
    not be necessary. Are there deterministic constructions for
    $(0,1)$-matrices with the property that their entry-wise product
    with a CS matrix improves CS performance? A natural class of
    candidates would be the incidence matrices of \hbox{$t$-$(v,k,\lambda)$}
    designs (see \cite{BJL} for example). Some related work is
    contained in \cite{mypaper-CScomp,mypaper-PBD}.
\end{question}

\section*{Acknowledgements}

The authors acknowledge the comments of anonymous referees.

The first and third authors have been supported by the Engineering
and Physical Sciences Research Council grant EP/K00946X/1.\\
The second author acknowledges the support of the Australian Research
Council via grant DP120103067, and Monash University where much of
this work was completed. This research was partially supported by the
Academy of Finland (grants \#276031, \#282938, and \#283262).

%\advance\bibitemsep by -0.5pt

\end{document}